\documentstyle[12pt,epsf]{article}

\title{Can conventional forces really explain the anomalous acceleration of
Pioneer 10/11 ?}

\author{J. P. Mbelek \\ Service d'Astrophysique, C.E. Saclay \\ F-91191
Gif-sur-Yvette Cedex, France, \\ \\ M. Michalski \\ Abt.
Quanteninformationsverarbeitung, Universit\"at Ulm \\ D-89069 Ulm, Germany}

\begin{document} \maketitle \baselineskip=8mm

\begin{abstract} A conventional explanation of the correlation between the
Pioneer 10/11 anomalous acceleration and spin-rate change is given.  First, the
rotational Doppler shift analysis is improved.  Finally, a relation between the
radio beam reaction force and the spin-rate change is established.  Computations
are found in good agreement with observational data.  The relevance of our
result to the main Pioneer 10/11 anomalous acceleration is emphasized.  Our
analysis leads us to conclude that the latter may not be merely artificial.
\end{abstract}

\section{Introduction} First published in 1998, results from an almost twenty
years study of radiometric data from Pioneer 10/11, Galileo and Ulysses
spacecraft have been continuously reported by Anderson et al.~\cite{Andersona}.
They indicate an apparent anomalous, constant, acceleration acting on the
spacecraft with magnitude $a_{P} = (8.74 \,\pm \,1.33)$ x $10^{-8}$ cm s$^{-2}$,
directed towards the Sun, to within the accuracy of the Pioneers' antennas.
Besides, an independent analysis of radio Doppler tracking data from the Pioneer
10 spacecraft for the time period 1987-1994 confirms the previous
observations~\cite{Markwardt}.  Also, the possibility that there exists an error
in the JPL's ODP program have been removed by using an independent program,
CHASMP.

Now, a number of potential causes have been ruled out by the Anderson {\sl et
al.}  team, namely gravity from Kuiper belt, gravity from the Galaxy, spacecraft
"gas leaks", anisotropic heat (coming from the RTGs) reflection off of the back
of the spacecraft high-gain antennae (Katz's proposal~\cite{Katz}, see Anderson
et al.~\cite{Andersonk}), radiation of the power of the main-bus electrical
systems from the rear of the craft (Murphy's proposal~\cite{Murphy}, see
Anderson et al.~\cite{Andersonm}), errors in the planetary ephemeris, and errors
in the accepted values of the Earth's orientation, precession, and nutation, as
well as nongravitational effects such as solar radiation pressure, precessional
attitude-control maneuvers and a possible nonisotropic thermal radiation due to
the Pu$^{238}$ radioactive thermal generators.  Indeed, according to the
authors, none of these effects may explain the apparent acceleration and some
are 3 orders of magnitude or more too small.  So, they conclude that there is an
unmodeled acceleration towards the Sun of ($8.09 \,\pm \,0.20)~10^{-8}$ cm
s$^{-2}$ for Pioneer 10 and ($8.56 \,\pm \,0.15)~10^{-8}$ cm s$^{-2}$ for
Pioneer 11.

In a further study Anderson {\sl et al.}~\cite{Andersonb}, observed that the
difference of the spin-rate history for the Pioneers 10 and 11 explains the
small difference of magnitude of the anomalous acceleration for Pioneer 10 and
for Pioneer 11.  The crucial point is that, removing the spin-rate change
contribution, one is left with an anomalous acceleration of the same amount with
a great accuracy (($7.84 \,\pm \,0.01)~10^{-8}$ cm s$^{-2}$ instead of ($8.74
\,\pm \,1.33)~10^{-8}$ cm s$^{-2}$) during a very long time interval (almost 20
years) for both Pioneer 10/11 to explain.  In view of the latter point, it is
clear that a conventional explanation of the Pioneer 10/11 anomalous
acceleration versus spin-rate change would at the same time clarify and
emphasize the possible importance of the main Pioneer anomaly for fundamental
physics.  Finally, the Pioneer 10/11 anomaly should deserve more serious
attention both on the theoretical and observational grounds.

\section{Study of the Pioneer 10/11 anomalous acceleration versus spin-rate
change}

\subsection{Reformulation of the new correlation} From the study of the Pioneer
10/11 anomalous acceleration, $a_{P}$, Anderson {\sl et al.}~\cite{Andersonb}
have found a correlation between $a_{P}$ and the rotational acceleration,
$\ddot{\theta}$, of both spinning spacecraft.  They expressed this as follows
\begin{equation} \label{aP vs spin-rate change} a_{P} = a_{P} (0) \,-
\,\kappa\,\ddot{\theta}, \end{equation} where $\kappa$ is a constant with unit
of length, $a_{P} = a_{P} (\ddot{\theta})$ and $a_{P} (0) \simeq 8$ x $10^{-8}$
cm~s$^{-2}$ are respectively the Pioneer anomalous acceleration with and without
any spin-rate change, and $\ddot{\theta}$ is the rotational acceleration derived
from the best fit to the data~\cite{Andersonb}.  The overall study is based on
the observation of discrepancies between the frequency of the re-transmitted
signal observed by the DSN (Deep Space Network) antennas, ${\nu}_{obs} (t)$, and
the predicted frequency of that signal, ${\nu}_{model} (t)$.  The observed
two-way anomalous effect is expressed (in the first order in $v/c$) as
\begin{equation} \label{anomalous effect} [ \,{\nu}_{obs} (t) \,-
\,{\nu}_{model} (t) \,]_{DNS} = \,- \,2 \,{\nu}_{0} \,\frac{a_{P}}{c} \,t,
\end{equation} where ${\nu}_{0}$ is the reference downlink frequency and
${\lambda}_{0} = c/{\nu}_{0} = 13.06$ cm is the corresponding wave-length.
Combining both relations (\ref{aP vs spin-rate change}) and (\ref{anomalous
effect}) yields \begin{equation} \label{anomalous effect reformulation} [
\,{\nu}_{obs} (t) \,- \,{\nu}_{model} (t) \,]_{DNS} = 2 \,{\nu}_{0}
\,\frac{\Delta v}{c} \,+ \,2 \,\frac{\kappa}{{\lambda}_{0}} \,\Delta
\dot{\theta}, \end{equation} where $\Delta v = - \,a_{P} (0)\,t$ and $\Delta
\dot{\theta} = \ddot{\theta} \,t$ denotes respectively the variations of the
radial velocity and the spin-rate of the spacecraft.  Clearly, relation
(\ref{anomalous effect reformulation}) above suggests that beside the familiar
Doppler effect connected to the linear velocity, a less commonly known frequency
shift that is connected to the rotational velocity may be acting.  The latter
effect is known indeed both on the theoretical~\cite{Bialynicki-Birula,
Courtiala, Mashhoon} and observational~\cite{Nienhuis, Courtialb, Garetz,
Bretenaker} grounds being referred to as the rotational Doppler effect (RDE).
According to the authors, the RDE due to the spin component of the beam has
already been taken into account in~\cite{Andersonb} (sec.  III-E) talking about
circular polarisation rather than a spin eigenstate.  Nevertheless, the authors
describe the part of the RDE due to spin, as circularly polarised
electromagnetic (EM) radiation has a spin of $\hbar$ per photon.  Now, a closer
look at their eq.  (15) reveals an inconsistency in their modelling of the RDE.

\subsection{The RDE contribution} Let us consider the influence of RDE on the
frequency received by DSN in detail.  In the first order in $v/c$ we have
\begin{equation}\label{eq_obsdsn} \nu_{observedDSN}=\left(1-\frac{v}{c} \right)
\left( \nu_{sentPioneer} \pm \nu_R \right), \end{equation} where
$\nu_R=\dot{\theta}/2 \pi$ is the rotation frequency of the Pioneer spacecraft,
and $\pm$ is for the sign of the downlink signal circular polarization.  The
frequency sent by Pioneer is first shifted by $\nu_R$ into the non-rotating, but
co-flying with the Pionner frame, and then shifted by linear Doppler effect.
Pioneer communication system converts the transmitted frequency:
\begin{equation} \nu_{sentPioneer}=f \nu_{observedPioneer}, \end{equation} where
$f = 240/221$ is the frequency turnaround ratio (see.~\cite{Andersonb}, sec.  II
D and sec.  III E).  Now, \begin{equation}\label{eq_obspioneer}
\nu_{observedPioneer}=\left(1-\frac{v}{c} \right) \nu_{sentDSN} \pm \nu_R,
\end{equation} where the frequency sent by DSN is first shifted by linear
Doppler effect in the co-flying, non-rotating frame, and then by RDE.  The $\pm$
sign is for the uplink signal circular polarization helicity.  Note that the
$\pm$ sign in eqs.  \ref{eq_obsdsn} and \ref{eq_obspioneer} could be different,
depending on down- and uplink signal polarizations, respectively.  Hence, in the
first order in $v/c$ we have \begin{equation} \nu_{observedDSN}=\left(1-\frac{2
v}{c} \right) \nu_0 \pm \left(1-\frac{v}{c} \right) \nu_R (1 \pm f),
\end{equation} and $\nu_0=f \nu_{sentDSN}$ is the downlink frequency.  As
Anderson {\sl et al.}  \cite{Andersonb} argue, that the rotation of the
spacecraft always \emph{increases} the radio frequency, we take $+$ sign instead
of both $\pm$'s.  As one can see, the additional term $( \,1 \, + \,f \,)
\,\left( \,1 \,- \,\frac{v}{c} \,\right) \,\nu_R$ that is involved by the RDE is
just missing in eq.  (15) of~\cite{Andersonb}.  The comparison with relation
(\ref{anomalous effect reformulation}) shows that this gives a contribution to
$\kappa$ approximately equal to $\frac{1 \,+ \,f}{2}
\,\frac{{\lambda}_{0}}{2\pi}$.  As this amounts to only about $1/15$ of the
whole, we still need to explain the remainder.  Besides, since the beam does not
possess helical phase fronts, the orbital angular momentum in the EM beam is
zero.  This leads us to search for a dynamical explanation of relation (\ref{aP
vs spin-rate change}) that may compete with the RDE (kinematical effect).

\subsection{The radio beam dynamical contribution} Radio beam reaction force
have been already discussed in~\cite{Andersonb} (sec.  VIII-A).  The authors
concluded that this would yield a substantial contribution $a_{rp} = ( \,1.10
\,\pm \,0.11 \,)$ x $10^{-8}$ cm~s$^{-2}$ to $a_{P}$ (according to us this
result should be doubled).  Also, the spin-rate change produced by the torque of
radiant power directed against the rotation have been investigated
in~\cite{Andersonb} (sec.  VIII-B).  However, in both cases the authors did not
try to formulate from the above considerations the possible link between $a_{P}$
and $\ddot{\theta}$.  Now, it is proved~\cite{Beth} that emitted or absorbed
photons can convert their angular momenta into a torque applied to a solid.  As
one knows, these photons may convert at the same time their momenta into a force
applied to a solid.  Let us emphasize that although the radiation pressure is
proportional to the emitted power, the torque is not.  Since the spacecraft
transmits continuously their signals to the Earth even without receiving any,
this provides a continuous contribution to the dynamical effect we are looking
for.  A rough estimate shows that this is by far the major contribution.  Let us
show how one can explain in this way the correlation between the different
values of $a_{P}$ and the spin-rate data of Pioneer 10/11 (whatever the value of
the torque).  Let $\ddot{\theta}_{0}$ be the spin-rate change of the spacecraft
due to other causes than the absorption or emission of EM radiation at the
communication frequencies.  Hence, there will be an additional contribution to
the spin-rate change related to the interaction with the photons, $\Delta
\ddot{\theta} = \ddot{\theta}(t) \,- \,\ddot{\theta}_{0}$, given by
\begin{equation} \label{Newton 2nd law for rotation} J \,\Delta \ddot{\theta} =
\Delta \dot{N} \,\hbar, \end{equation} where $J$ is the moment of inertia along
the spin axis of the spacecraft and $\Delta \dot{N}$ denotes the number
difference between absorbed and emitted photons per unit of time.  Recall that
the angular momentum of a photon of an EM wave is $\pm \hbar$ with the sign
depending on the helicity of the EM wave, namely right or left circularly
polarized.  Also, a photon of frequency $\nu = \omega/2\pi$ carries a momentum
$\hbar \omega/c$.  Moreover, a photon emitted at an angle $\theta$ with respect
to the spin axis transfers to the spacecraft a momentum $( \,\hbar \omega/c \,)
\,\left( \,1 \,+ \,\cos{\theta} \,\right)$ after reflexion on the parabolic
dish.  As, the computation made in~\cite{Andersonb} (eq.  (37), sec.  VIII-B)
manifestly only accounts for the case $\theta = 0$ that is a momentum transfer
$2 \,\hbar \omega/c$ per photon, substracting the latter yields a remainder of
momentum transfer to the spacecraft the magnitude of which amounts to $p(\theta)
= ( \,\hbar \omega/c \,) \,\left( \,1 \,- \,\cos{\theta} \,\right)$ per photon
emitted at an angle $\theta$.  So, the spacecraft will be subject to an
acceleration in excess $\Delta a_{P}$ given by \begin{equation} \label{Newton
2nd law for translation} M \,\Delta a_{P} = \int_{0}^{{\Omega}_{beam}} \,\left(
\,p(\theta) \,+ \,p(-\,\theta) \,\right) \,\frac{\Delta
\dot{N}}{{\Omega}_{beam}} \,d\Omega = \Delta \dot{N} \,\frac{\hbar \omega}{c}
\,\frac{{\Omega}_{beam}}{2\pi}, \end{equation} where $M$ is the mass of the
spacecraft, $\Omega = 2\pi \,\left( \,1 \,- \,\cos{\theta} \,\right)$ the solid
angle subtended by the cone with half-aperture $\theta$ and ${\Omega}_{beam}$
denotes the total solid angle of the conical beam within which the photons are
emitted.  Because of the way the high-gain antenna of the spacecraft is directed
with respect to the Earth, clearly $\Delta a_{P}$ is in the opposite direction
to $a_{P}(0)$ and consequently substracts to it.  Thus, one gets a net apparent
anomalous acceleration $a_{P} = a_{P}(\ddot{\theta}_{0}) \,-\,\Delta a_{P}$
which yields combining relations (\ref{Newton 2nd law for rotation}) and
(\ref{Newton 2nd law for translation}) \begin{equation} \label{Delta aP vs
spin-rate change} a_{P} = a_{P} (0) \,- \,{\Omega}_{beam} \,\frac{\nu}{c}
\,\frac{J}{M} \,\ddot{\theta}.  \end{equation}

\subsection{Both contributions} So, the comparison of relation (\ref{Delta aP vs
spin-rate change}) above with (\ref{aP vs spin-rate change}) yields on account
of the RDE term \begin{equation} \label{kappa} \kappa = \frac{1}{{\lambda}_{0}}
\,\frac{J}{M} \,{\Omega}_{beam} \,+ \,\frac{1 \,+ \,f}{2}
\,\frac{{\lambda}_{0}}{2\pi} \simeq \frac{2\pi}{{\lambda}_{0}} \,\frac{J}{M} \,(
\,1 \,- \,\cos{\Theta} \,) \,+ \,\frac{1 \,+ \,f}{2}
\,\frac{{\lambda}_{0}}{2\pi}, \end{equation} where $\Theta$ denotes the angle
between the antenna axis (spin axis) and the first diffraction minimum of the
conical beam.  The nominal working values of Pioneer 10 used by Anderson {\sl et
al.}  are $M = 251.883$~kg, $J \approx 588.3$~kg~m$^{2}$ and $\Theta \approx
4^{\circ}$ (see~\cite{Andersonb}, sec.  VIII-A).  These yield $\kappa \simeq 30$
cm in good agreement with $\kappa = 30.7 \pm 0.6$ cm found in~\cite{Andersonb}.
Thus, it seems that one aspect of the Pioneer anomaly has found a conventional
explanation which can be checked in the laboratory.  As regards the constant
part $a_{P} (0)$ of the anomalous acceleration (the main Pioneer 10/11 anomalous
acceleration), it still remains to explain.\\

\section{Conclusion} We have provided a conventional and quantitative
explanation to the small difference of magnitude of the anomalous acceleration
observed between the Pioneers 10 and 11.  As observed by Anderson {\sl et
al.}~\cite{Andersonb}, we find that this is related to the difference of
spin-rate history for both spacecraft.  Now, these authors have shown that when
the spin-rate change contribution is removed, one is left with an anomalous
acceleration $a_{P} (0)$ of the same amount with a great accuracy during a very
long time interval for both Pioneer 10/11 to explain.  Since the Pioneers 10 and
11 are remote spacecraft moving in opposit directions from the Sun, this makes
difficult to understand such a persistant constant anomalous acceleration in
terms of conventional physics.  Hence, the conventional explanation we suggest
for the Pioneer 10/11 anomalous acceleration versus spin-rate change points out
in favour of the possibility that something new of physical interest may be
responsible of the main Pioneer anomaly at the expense of any possible internal
cause (see~\cite{Scheffer}, for the revival of such a possibility).

{\large {\bf acknowledgments}}

The authors are grateful to Johannes Courtial for all his helpful comments and
remarks in preparing this paper.

\end{document}